\documentclass[twocolumn]{aa}
\usepackage{graphicx}
\usepackage{txfonts}
\begin{document}

\title{An extensive library of 2500-10500 \AA\ synthetic 
       spectra\thanks{Figure~1 available only in electronic form at CDS 
       via anonymous ftp to cdsarc.u-strasbg.fr (130.79.128.5) or via
       http://cdsweb.u-strasbg.fr/cgi-bin/qcat?J/A+A/.
       The entire library of synthetic spectra is accessible via
       http://archives.pd.astro.it/2500-10500/ or (for the 1~\AA/pix version only)
       http://gaia.esa.int/spectralib/spectralib1A/ SpectraLib1a.cfm}
      }

   \author{U. Munari\inst{1},  
          R. Sordo\inst{1,2},
          F. Castelli\inst{3}
          \and
          T. Zwitter\inst{4}
          }

   \offprints{munari@pd.astro.it}

   \institute{INAF Osservatorio Astronomico di Padova, via dell'Osservatorio 8, 36012 Asiago (VI), Italy
         \and
              Dipartimento di Astronomia, Univ. di Padova, vicolo della Specola 5, 35122 Padova, Italy
         \and 
              INAF - Osservatorio Astronomico di Trieste, Via G.B. Tiepolo 11, 34131 Trieste, Italy
         \and
              University of Ljubljana, Department of Physics, Jadranska 19, 1000 Ljubljana, Slovenia            
              }

   \date{Received YYY ZZ, 2005; accepted YYY ZZ, 2005}

    \abstract{We present a complete library of synthetic spectra based on
    Kurucz's codes that covers the 2500--10\,500 \AA\ wavelength range at
    resolving powers $R_{\rm{P}}$=20\,000, 11\,500 ($\equiv$ GAIA), 8500
    ($\equiv$ RAVE), 2000 ($\equiv$ SLOAN) and uniform dispersions of 1 and
    10~\AA/pix. The library maps the whole HR diagram, exploring 51\,288
    combinations of atmospheric parameters spanning the ranges: 3500 $\leq
    T_{\rm eff} \leq 47\,500$ K, 0.0$\leq \log g \leq 5.0 $, $-$2.5 $\leq$
    [M/H] $\leq$ 0.5, [$\alpha$/Fe] = 0.0,+0.4, $\xi$ =1,2,4~km sec$^{-1}$, 0
    $ \leq V_{\rm rot} \leq$ 500~km sec$^{-1}$. The spectra are available both
    as absolute fluxes as well as continuum normalized.  Foreseen
    applications of the library are briefly discussed, including automatic
    classification of data from spectroscopic surveys (like RAVE, SLOAN,
    GAIA) and calibration of differential photometric indexes. Data
    distribution and access to the library via dedicated web page are
    outlined.

              }

   \maketitle

\section{Introduction}

The interest in large and complete synthetic spectral libraries of normal
stars ranges over many different application areas, including ($a$)
automatic analysis and classification of large volumes of data, like those
collected by ongoing spectral surveys, ($b$) derivation of radial velocities
via cross-correlation against best matching templates, ($c$) calibration of
spectroscopic line/band classification criteria, and ($d$) calibration of
photometric indexes. Tasks $a$-$c$ require the synthetic spectra to match
the absorption lines in observed spectra normalized to the local continuum,
generally irrespective of how well the overall spectral energy distribution
is reproduced. The goal of task $d$ is instead to match the energy contained
within photometric bands, irrespective of the performance on the individual
absorption lines. Therefore, synthetic spectral libraries well suited for
one task does not necessarily perform equally well on another. For example,
within the Kurucz's synthetic spectra suite, the inclusion in the
computations of the so called `predicted lines' (hereafter PLs, Kurucz 1994)
limits the range of applications. The PLs are calculated absorption lines
due to transitions involving one and sometimes two levels whose locations
are predicted by atomic structure codes. Wavelengths for these lines may be
uncertain by $\geq$10~\AA, with large uncertainties affecting also their
computed intensity. As a consequence the predicted lines may not correspond
in position and size to the observable counterparts (cf. Bell et al. 1994,
Castelli and Kurucz 2004). For all these reasons the predicted lines are not
intended to be used for the analysis of high resolution spectra, but are an
essential contribution to the total line blanketing in model atmospheres and
spectral energy distribution computations. Very accurate line wavelengths
and shapes are in fact not needed for the statistical computations of the
line opacity in model atmospheres as well as for the prediction of low
resolution spectra.

Various libraries of synthetic spectra are becoming available, and they are
summarized and compared in their basic characteristics in Table~1. These
libraries widely differ in wavelength interval, spectral resolution, input
model atmospheres and range of atmospheric parameters explored, so that
their overlapping is not significant.

As other libraries in Table~1, our one is based on Kurucz's atmospheres,
line-lists and computing software. The main goal of our spectral library are
spectroscopic applications of the $a$ and $b$ types above, which imposes for
ex. the exclusion of the PLs from computations and therefore no pretension
of accurate broad-band photometric performances particularly at the lower
temperatures. Other relevant characteristics of our library are the various
resolutions and rotational velocities at which it is provided, its wide
wavelength range, the adoption of improved model atmospheres based on the
new opacity distribution functions (ODFs) by Castelli and Kurucz (2003), the
use of the TiO line list of Schwenke (1998), the inclusion of
$\alpha$-element enhancement and different micro-turbulent velocities.

  \begin{figure}[!t]
     \centering
     \caption{Coverage of the $\log g$, $T_{\rm eff}$ plane by the
      synthetic spectral grid presented in this paper (figure available electronic only).
      {\em Panel a}: spectra with [$\alpha$/Fe]=0.0 and $\xi$=2~km~sec$^{-1}$; 
      {\em Panel b}: spectra with [$\alpha$/Fe]=+0.4 and $\xi$=2~km~sec$^{-1}$;
      {\em Panel c}: spectra with [$\alpha$/Fe]=+0.4 and $\xi$$\neq$2~km~sec$^{-1}$}
     \label{grid}
  \end{figure}

\section{The spectra}

The whole grid of spectra in our library was computed using the SYNTHE code
by Kurucz (Kurucz and Avrett 1981, Kurucz 1993), running under VMS operating
system on a Digital Alpha workstation in Asiago. We adopted as input model
atmospheres the ODFNEW models
(http://wwwuser.oat.ts.astro.it/castelli/grids/; Castelli and Kurucz 2003).
They differ from the NOVER models (http://kurucz.harvard.edu/grids.html;
Castelli et al. 1997) for the adoption of new ODFs, replacement of the solar
abundances from Anders and Grevesse (1989) with those from Grevesse and Sauval
(1998), and improvements in the molecular opacities among which the adoption
of the molecular line-lists of TiO by Schwenke (1998, as distributed by
Kurucz 1999a) and of H$_{2}$O by Partridge and Schwenke (1997, as distributed
by Kurucz 1999b). For the combination of atmospheric parameters for which no
ODFNEW models were available at the time of writing, we adopted the
corresponding NOVER input model atmospheres. We plan to update the library
from NOVER to ODFNEW models as more of the latter will become available and
to post the result on the library web page (see below).  We remark that both
the NOVER and the ODFNEW models were computed with the overshooting option
for the mixing-length convection switched off, while Kurucz (K) atmospheric
models (also available from http://kurucz.harvard.edu/grids.html) were
computed with the overshooting option switched on. Several papers have
demonstrated that for stars with active convection ($T_{\rm eff}$$<$9000 K)
the no-overshooting convection treatment provides better agreement with the
observations than the overshooting case does (Castelli et al. 1997, Smalley
and Kupka 1997, Gardiner et al. 1999, Smalley et al. 2002). The
no-overshooting models used by us were computed for the mixing-length
parameter to the scale height of 1.25. This value allows to fit the observed
solar irradiance, while a lower value, like that of 0.5 suggested by Smalley
et al., 2002, does it not. On the contrary, 0.5 seems to better fit the
wings of H$\beta$, provided that the position of the continuum is known with
an uncertainty smaller than 1\%. In fact, a difference of 1\% in the
position of the solar continuum corresponds to the difference between 0.5
and 1.25 of the mixing-length parameter (Castelli et al., 1997). Generally,
it's very difficult to state the location of the continuum across the wings
of the Balmer lines, especially in Echelle spectra which are notoriously
severely affected by the blaze function.

  \begin{figure}
     \centering
     \resizebox{\hsize}{!}{\includegraphics{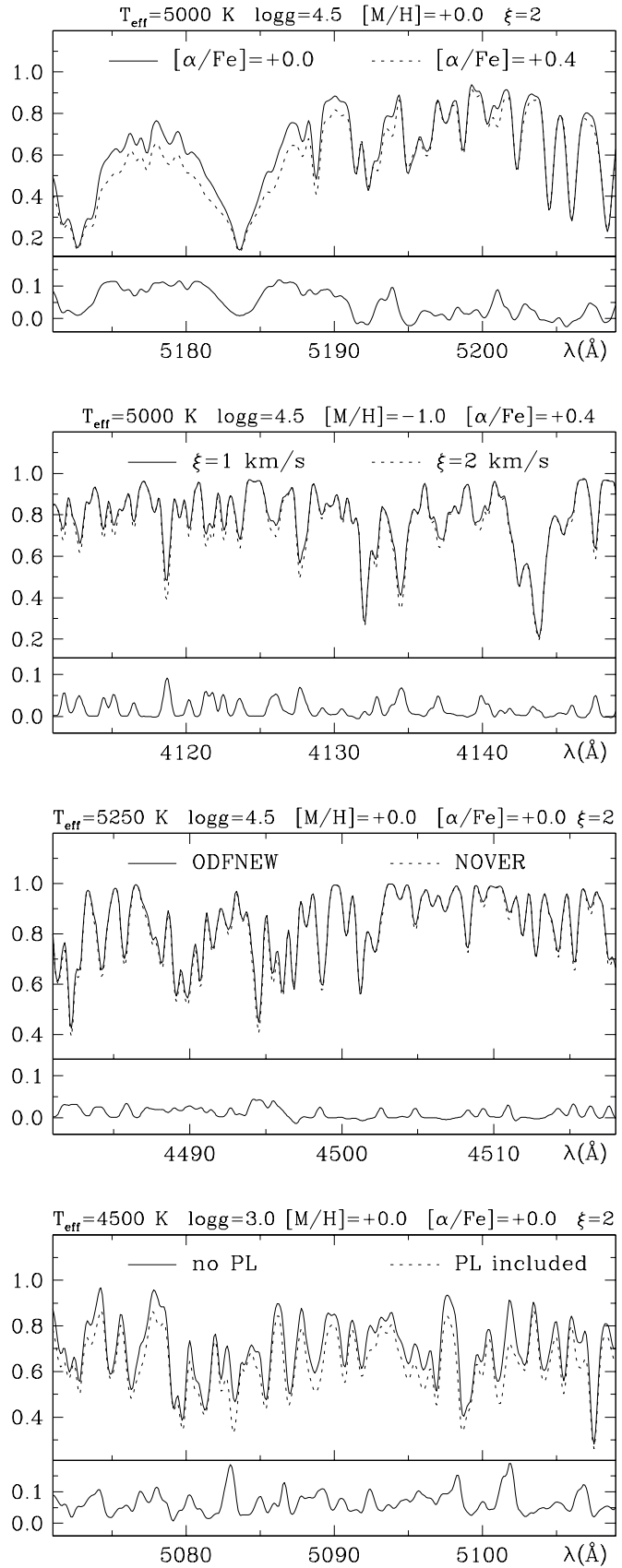}}
     \caption{The effect of $\alpha$$-$enhancement, micro-turbulent velocity
     $\xi$, new vs old ODFs, and inclusion of `predicted lines' are
     illustrated by sample spectra. To increase readability all spectra have
     been broadened to a rotational velocity of 20 km~sec$^{-1}$. In each
     panel the bottom part highlight the difference between the plotted
     spectra (always in the sense {\em solid} $-$ {\em dotted} lines).}
     \label{confronto}
  \end{figure}

  \begin{table*}
   \centering
   \caption[]{Comparison of the principal characteristics of publicly
              available libraries of synthetic spectra with resolution
              better than 5~\AA.  {\it N}: number of combinations of
              atmospheric parameters and rotational velocities.  {\it new
              ODF}: use of the new opacity distribution functions.  {\it
              TiO}: inclusion of the TiO line-list. {\it pred. lines}:
              inclusion of the ` predicted lines'. The last three columns
              indicate if the spectra are provided for different values of
              enhancement of $\alpha$-elements, micro-turbulence velocity
              and rotational velocity.}
   \includegraphics[width=17.5cm]{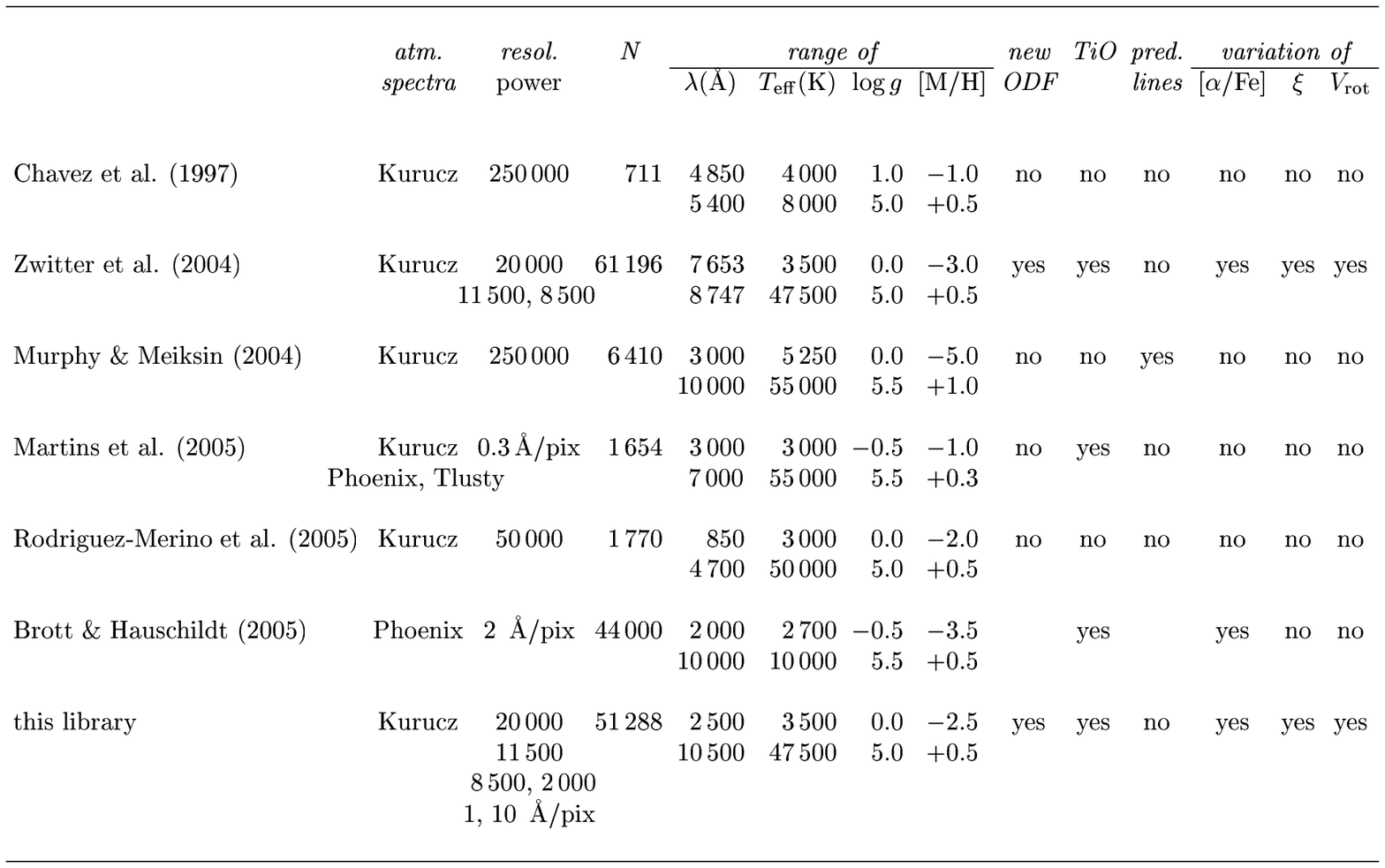}  
   \label{survey}
  \end{table*}
  
  \begin{table*}
   \caption[]{Range of atmospheric parameters explored by our synthetic spectral library.}
   \label{Step}
   \begin{center}
   \begin{tabular}{lll}
  \hline
  \\
  temperature (K)       	&~3500 $\leq$~~~~$T_{\rm eff}$~~$\leq$ 47\,500	
                                &steps of 250 K for $T_{\rm eff} \leq $10\,000 K; 
                                 proportionally larger for higher $T_{\rm eff}$\\ 
  gravity (cgs units)   	&~~~~0.0 $\leq$~~$\log{g}$~~$\leq$ 5.0 	&\\
  metallicity           	&\,\,$-$2.5 $\leq$ [M/H] $\leq$ +0.5       &\\
  rotation velocity (km sec$^{-1}$)   &~~~~~~~0 $\leq$ ~~{\sl V}$_{\rm rot}$ 
                                ~~$\leq$ 500 & 0,2,5,10,15,20,30,40,50,75,100 
                                \hfill{for $T_{\rm eff} \leq $ 6000 K}\\
                                &&0,10,20,30,40,50,75,100,150,200,250,300,400,500 
                                 \hfill{for  $T_{\rm eff} >$ 6000 K}\\
  enhancement            	& ~~~~~~[$\alpha$/Fe] = 0.0, +0.4             &
                                [$\alpha$/Fe]=+0.4 for [M/H]=$-$0.5, $-$1.0, $-$1.5\\
  micro-turbulent velocity (km sec$^{-1}$) &\multicolumn{1}{c}{$\xi$ = 1, 2, 4}
                                & 1 and 4 km sec$^{-1}$ for [$\alpha$/Fe]=+0.4\\
  ODFs                          &\multicolumn{1}{c}{new}&
                                old where new ones not yet available\\
  predicted lines               & \multicolumn{1}{c}{excluded} \\
  \hline
  \end{tabular}
  \end{center}
  \end{table*}

In computing the synthetic spectra we used the Kurucz's atomic and molecular
line-lists (Kurucz 1992). The effects of several molecules were taken into
account, including C$_2$, CN, CO, CH, NH, SiH, SiO, MgH and OH. The TiO
molecular lines (from Schwenke 1998) were included in the computation when
$T_{\rm eff}\leq$~5000~K. The synthetic spectra are calculated and made
available both as fluxes for unit wavelength and as normalized intensities.
The latter have a different meaning from the normalization to the continuum
of the observational spectroscopy jargon, where the location of the
continuum is taken to be represented by the regions away from absorption
lines or bands. In synthetic spectra, a continuum normalized spectrum is
obtained by dividing the absolute fluxed spectrum by its {\em calculated}
continuum. Noticeable differences are essentially limited to the head of the
Paschen and Balmer series of hydrogen, and to the coolest stars dominated by
molecular bands.

\section{The library}

The spectra in our library covers from 2500 to 10500 \AA\ and were
calculated at a resolving power R$_{\rm P}$=$\lambda/\Delta
\lambda$=500\,000 and then degraded by Gaussian convolution to lower
resolving powers and properly re-sampled to Nyquist criterion (the FWHM of
the PSF being 2 pixels), to limit the data volume and therefore facilitate
the distribution.

The highest resolving power for our library is $R_{\rm P}$=20\,000, which is
typical of moderately high resolution Echelle observations, as the spectral
database of real stars of Munari and Tomasella (1999) and Marrese et al.
(2003). The library is also provided at the resolving powers of the SLOAN
survey $R_{\rm P}$=2\,000 (York et al. 2000), of the RAVE survey $R_{\rm
P}$=8\,500 (Steinmetz 2003, Munari et al. 2005) and of the Gaia space
mission $R_{\rm P}$=11\,500 (Katz et al. 2004, Wilkinson et al. 2005).
Furthermore, the present library is also made available at the uniform
dispersions of 1 and 10~\AA/pix (again with the FWHM of the PSF
being 2 pixels). The latter two are intended to be of assistance in
classification of medium and low resolution grating spectra obtained in
single dispersion mode (as with conventional B\&C spectrographs or
EFOSC-like imager/spectrographs).

Each spectrum is provided for a range of rotational velocities, as detailed
in Table~1, spanning 14 values between 0 and 500~km~sec$^{-1}$ for stars
hotter than 6000~K and 11 values between 0 and 100~km~sec$^{-1}$ for cooler
ones.

The explored combinations of parameters in Table~2 give rise to 51\,228
spectra, each provided in six different combinations of resolving power and
sampling, and two flux varieties (absolute units and normalized
intensities). The total number of individual spectra contained in this
library is therefore 51\,288$\times$6$\times$2=615\,456. The spectra are
provided as gzipped ASCII files containing only the flux column, the common
wavelength column being given separately only once for each of the six
resolutions. The adopted scheme for the naming of files is outlined in
Table~3.

All spectra in the library can be directly accessed and retrieved through
the dedicated web page http://archives.pd.astro.it/2500-10500/. The version
of the library at 1~\AA/pix is accessible also via ESA's web site
http://gaia.esa.int/spectralib/, where browsing facilities based on Virtual
Observatory tools are provided. A distribution via DVDs will be possible in
special cases (to be arranged directly at munari@pd.astro.it).

  \begin{table}
  \caption[]{Naming scheme for the spectra contained in the library}
  \label{names}
  \begin{center}
  \begin{tabular}{c l l}
  \hline
  &&\\
  \multicolumn{3}{c}{example: ~~T04000G45M05V015K2ANWNVR20F.asc}\\
  &&\\
  {\em characters} &         & {\em meaning}    \\                    
  &&\\
    1--6      & T04000  & $T_{\rm eff}$ (K)\\
    7--9      &  G45    & 10 $\times \log{g}$ (cgs units)\\
     10       &   M     & sign for metallicity: M=`$-$', P=`+'\\
   11--12     &   05    & 10 $\times$ [M/H]\\
   13--16     &  V015   & rotational velocity (km sec$^{-1}$)\\
   17--18     &   K2    & microturbulent velocity (km sec$^{-1}$)\\
     19       &   A     & [$\alpha$/Fe] enhancement: S=0.0, A=+0.4 \\
   20--21     &   NW    & NW = new ODF models, \\
              &         & OD = old ODF models\\
   22--23     &   NV    & no overshooting\\
   24--26     &   R20   & R20 = resolving power 20\,000;\\
              &         & RVS = resolving power 11\,500 (GAIA);\\
              &         & RAV = resolving power 8500 (RAVE);\\
              &         & SLN = resolving power 2000 (SLOAN);\\ 
              &         & D01 = 1 \AA/pix dispersion; \\
              &         & D10 = 10 \AA/pix dispersion\\ 
     27       &   F     & F=fluxed spectrum (erg  cm$^{-2}$ sec$^{-1}$ \AA$^{-1}$ )\\
              &         & N=normalized spectrum\\
  &&\\
  \hline
  \end{tabular}
  \end{center}
  \end{table}

\section{Test applications}

To evaluate the spectroscopic and photometric performances of our library we
have carried out several tests. They support the fully satisfactory use of
the library in spectroscopic applications (our main goal), while caution
applies on the photometric use at the lower temperatures and/or shorter
wavelengths. It is worth noticing that significant discrepancies affect the
comparison of available libraries with observational data over blue
wavelengths and low temperatures.

\subsection{Eclipsing binaries}

A heavily demanding spectroscopic application of our library has been
carried out by Siviero et al. (2004) and Marrese et al. (2005) in
state-of-the-art investigation of double-lined solar type eclipsing
binaries, as a source for ($i$) best matching templates in cross-correlation
determination of radial velocities and ($ii$) reference grid in $\chi^2$
analysis of atmospheric properties of binary components. They used the
$R_{\rm P}$=20\,000 version of the library. 

  \begin{figure}
     \centering
     \resizebox{\hsize}{!}{\includegraphics{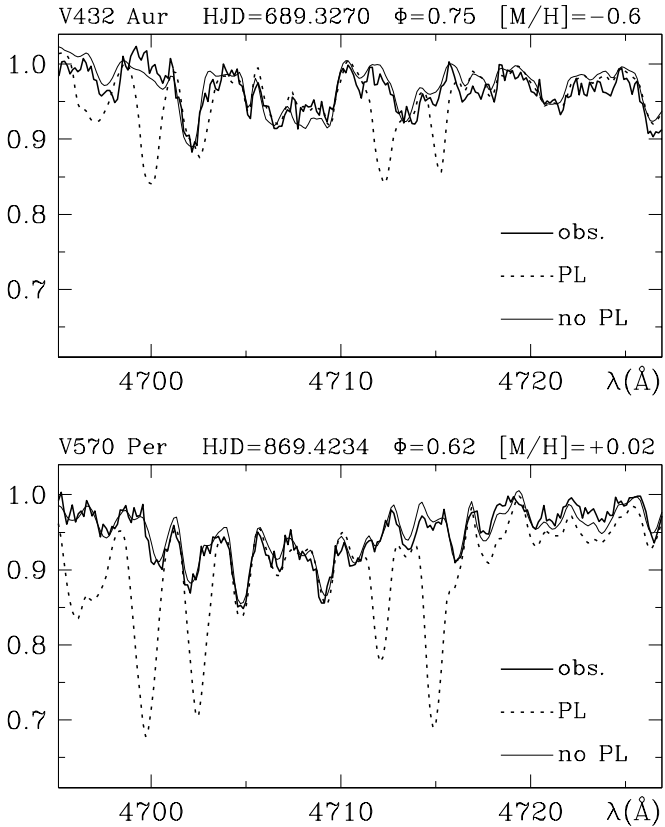}}
     \caption{The observed spectra (solid thick lines) of two eclipsing
     binaries are compared with synthetic ones computed without (solid thin
     line) and with (dotted line) inclusion of so called `predicted
     lines'. The synthetic spectra are computed for the atmospheric
     parameters ($T_{\rm eff}$, $\log g$) and co-rotational rotational
     velocity as derived from the accurate orbital solutions of Siviero et
     al. (2004, for V432 Aur) and Marrese et al. (2005, for V570 Per) and
     their chemical abundance analysis.  The spectra of the two components
     are shifted at the appropriate radial velocity and combined together
     following the light-ratio at the these wavelengths as derived in the
     orbital solution. It is evident from the comparison how the synthetic
     spectra computed {\em without} the `predicted lines' offer a much better
     representation of the observed ones, and how the effect of the
     inclusion of `predicted lines' get worst with increasing metallicity.}
     \label{binaries}
  \end{figure}

The availability of a wide and complete range of rotational velocities for
all the spectra in our library always allowed to select the best template
even with binaries characterized by high rotations, a key ingredient to
reach absolute precisions better than 1~km~sec$^{-1}$ in individual radial
velocities even on modest S/N spectra.

The derivation of atmospheric parameters via $\chi^2$ fitting to the library
converged independently to the same values as derived geometrically by the
orbital solution. The $\chi^2$ and orbital determinations provide the same
gravities within 0.03 dex, the same rotational velocities within
4~km~sec$^{-1}$ and the same difference in temperature between the two
components within 42~K (0.6\%). The metallicity was indirectly derived by
the orbital solution via matching with theoretical isochrones tracks on the
temperature-luminosity plane. The values derived in this way are confirmed
within 0.06 dex by direct $\chi^2$ analysis. Given the significant rotation
exhibited by components of binaries (generally synchronous with orbital
motion) the wide range of rotational velocities covered by our library
proved once again to be a mandatory ingredient for accurate $\chi^2$
results.

The above results would have been significantly degraded if, instead of our
one, it would have been used a library computed with inclusion of the PLs.
To support this assertion we have re-run exactly the same $\chi^2$ fits of
Siviero et al. (2004) and Marrese et al. (2005) using an identical grid of
synthetic spectra, this time however computed on purpose with the inclusion
of PLs. In the case of V570~Per the use of spectra with PLs pushes the
$\chi^2$ fit away from the geometrical solution by $\Delta$$T_{\rm
eff}$=109~K and $\Delta$[M/H]=0.16, and by $\Delta$$T_{\rm eff}$=98~K,
$\Delta$$\log g$=0.10 and $\Delta$[M/H]=0.11 for V432~Aur
(average values for the components of the given binary). The differences
are always in the sense that inclusion of PLs pushes the $\chi^2$ fit away
from the correct one by going toward cooler temperatures and lower
metallicities. This is an expected, systematic trend. In fact, the inclusion
of PLs causes the continuum in synthetic spectra to become fainter (cf.
bottom panel of Figure~2), causing a reduction in the contrast between strong
absorption lines (those dominating the outcome of the $\chi^2$ fit) and the
adjacent continuum. To re-establish the proper contrast, the $\chi^2$ fits a
cooler temperature. Also fitting a lower metallicity helps in re-establish
the proper contrast. In fact, PLs are mainly a forest of weak, optically
thin lines that respond much faster to a small change in metallicity than
the stronger absorption lines which are much closer to optically thick
conditions than PLs. 

We observed also a reduction in the accuracy of the radial velocities, which
is however more difficult to quantify. In fact, it appears as a systematic
shift (toward shorter or longer wavelengths) dependent on the actual
wavelength interval and the orbital phase. This is caused by the beating in
the cross-correlation between the pattern of PLs and the observed absorption
lines which is function of both $\lambda$ and radial velocity separation
between the binary components. The presence of these locally different
systematic shifts become evident only when radial velocities from various
wavelength intervals on the same spectrum are inter-compared (like those
provided by individual orders in an Echelle spectrum). The overall effect is
an increase of the scatter of individual measurements, different in amount from
spectrum to spectrum.

This concerns the analysis performed on wavelength regions dominated by weak
PLs, as in the example on the bottom panel of Figure~2. There are however
limited wavelength intervals where PLs much stronger than usual are
stochastically clustered together. The result of a $\chi^2$ fit performed on
any such interval would provide non-sense results. One such interval is
illustrated in Figure~3, where the observed spectra of V432~Aur and V570~Per
are taken from Siviero et al. (2004) and Marrese et al. (2005). While the
fit with synthetic spectra computed without PLs provide - for the values of the
atmospheric parameters derived by the geometrical orbital solution - an
almost perfect match to the observed spectra, the same spectra computed with
inclusion of PLs bear no resemblance at all to the observed spectrum. It is
evident that the photometric calibration of a color-index that includes a
narrow band centered on such a wavelength interval would provide non-sense 
results if carried out on synthetic libraries computed with PLs.

  \begin{figure*}
     \centering
	\includegraphics[width=12.0cm,angle=270]{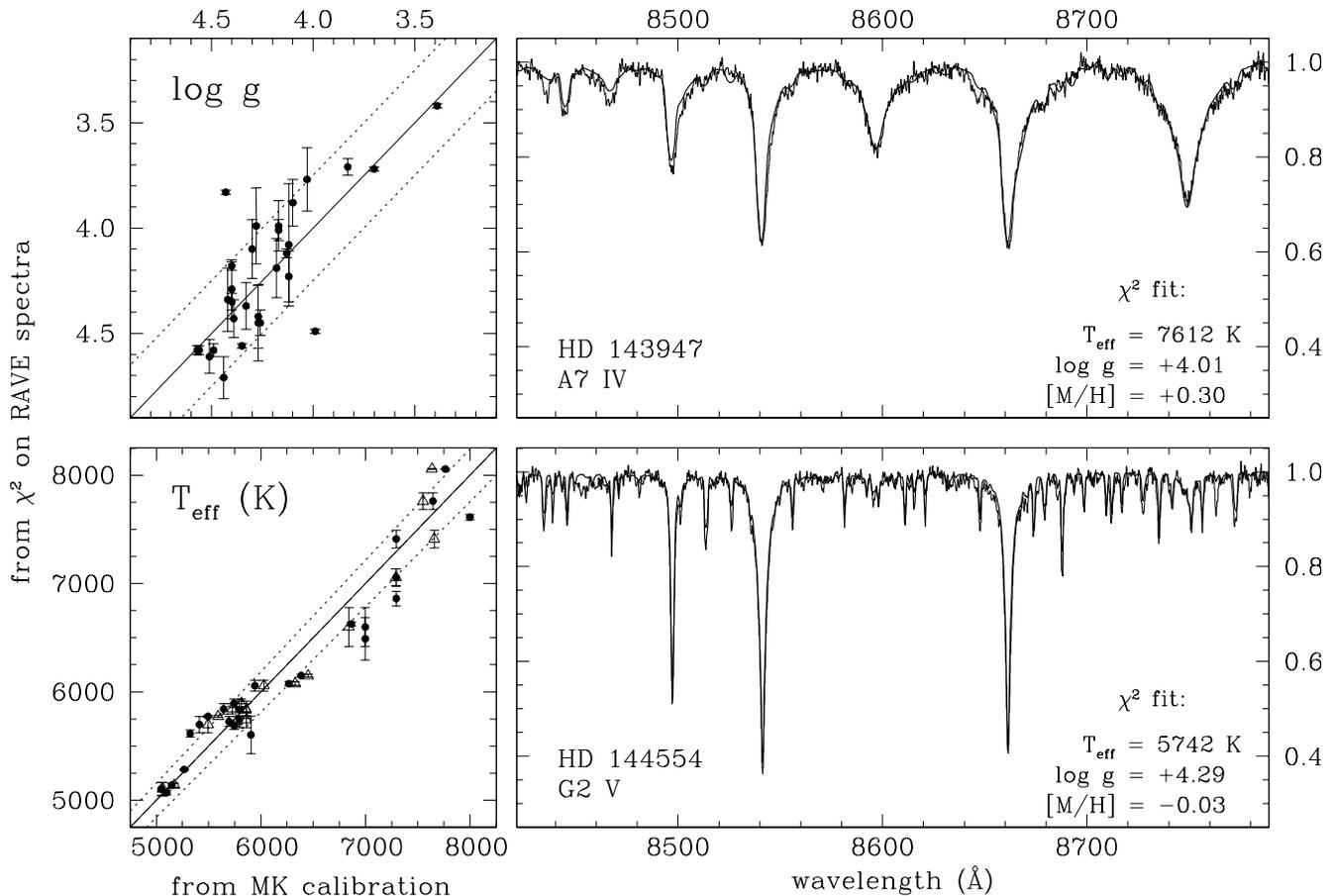}
     \caption{The figure illustrates the use of our synthetic spectral
     library at resolving power $R_{\rm P}$=8500 to classify RAVE spectra.
     The left panels show the results of $\chi^2$ fitting in gravity and
     temperature of spectra of HD stars from RAVE fields 1607m492 and
     1716m425, with spectral types from the Michigan Project (Houk 1978)
     and calibration into $T_{\rm eff}$ and $\log g$ by Straizys and
     Kuriliene (1981). The dotted lines represent margins for $\pm$3\% in
     $T_{\rm eff}$ and $\pm$0.25 dex in gravity. The error bars on each
     point represent the spread in the results of $\chi^2$ fitting of three
     different spectra of the same object secured in different nights. On
     the right panels, examples of match between observed and synthetic
     spectra are provided for a couple of HD stars from the same RAVE
     fields.  The results of the $\chi^2$ fitting for the two spectra are
     given for $T_{\rm eff}$, $\log g$, and metallicity.}
     \label{rave}
  \end{figure*}

\subsection{RAVE spectra}

We are engaged in the data analysis of the RAVE spectroscopic survey of the
galactic Halo (Steinmetz 2003, Munari et al. 2005) with the 6dF fiber
positioner and spectrograph at the UK-Schmidt at the Anglo-Australian
Observatory. The survey is carried out at high S/N over the 8420--8780 \AA\
interval at $R_{\rm P}$=8\,500 resolving power.

Two low galactic latitude calibration fields rich in HD stars (1607m492 and
1716m425) have been observed a few times during the survey. Spectral
classification of these stars are taken from the Michigan Project (Houk
1978). To test the performance of our library, we have $\chi^2$ fitted the
spectra of these HD stars against our library of synthetic spectra in its
$R_{\rm P}$=8\,500 version.  The results of the $\chi^2$ fitting are
presented in Figure~4. The right panels illustrate the accuracy of the
achieved fits by over-plotting observed and synthetic spectra for the sample
cases of a Sun analogue and of a A-type star. The $T_{\rm eff}$ and $\log g$
values derived by the $\chi^2$ fit well match those expected from the G2V
and A7IV classifications following the calibration by Straizys and Kuriliene
(1981). For the whole sample of HD stars observed in the two calibration
fields, the left panels of Figure~4 compare the $T_{\rm eff}$ and $\log g$
values derived by the $\chi^2$ fitting with those expected on the base of
the spectral classification and the Straizys and Kuriliene (1981)
calibration. The results in Figure~4 show an agreement within 3\% for
$T_{\rm eff}$ and better than 0.25 dex in $\log g$, the dispersion being
almost entirely accountable for by the natural width in both $T_{\rm eff}$
and $\log g$ of a spectral sub-type and a luminosity class in the MK
classification scheme.

\subsection{The distance to the Pleiades}

The present library, in the 1~\AA/pix version, was used in differential
broad-band photometric applications by Munari et al. (2004) while deriving a
precise distance to the Pleiades via orbital solution of the eclipsing
binary member star HD~23642, which spectral type is A2V.

Magnitudes in ten different photometric systems were found in literature for
this star, and they were used to derive temperature, gravity and reddening
of both components of the binary via calibration on our spectral libraries
of the color indexes of each photometric system. The zero point of colors
was fixed to 0.00 for a Vega model characterized by $T_{\rm eff}$=9400,
$\log g$=3.9, [M/H]=$-$0.5 and $\xi$=0.0~km~sec$^{-1}$ downloaded from
Kurucz's web site.
 
The photometric results on atmospheric and reddening parameters so derived
for both components of HD~23642 were confirmed by high resolution
spectroscopy and the orbital solution. The successful photometric
application of the library was made possible by the proximity in the
parameter space of HD~23642 and the calibrator Vega.

\subsection{Broad-band photometric tests}

To test the broad-band photometric performances of our library we have
investigated the reproduction of observed colors along the Main Sequence
(MS). The results are shown in Figures~5 and 6, where the $T_{\rm eff}$ and
$\log g$ of stars along the main sequence were adopted from Straizys and
Kuriliene (1981). We stopped at M2V as the MS faint end, because later types
are characterized by $T_{\rm eff}$ cooler than the 3500~K limit of our
library (cf. Table~4). We included as a term of comparison all the other
libraries from Table~1 for which data are actually accessible and are
characterized by a wavelength coverage wide enough to cover at least three
consecutive photometric bands. In this regard, a word of caution is
necessary when considering the results from Martins et al. (2005) library,
which red limit is placed at 7000~\AA\ where the $V$ band has still 1\%
transmission and does reach 0.0 only at 7400 \AA. The effect of missing this
far red wing of the $V$ transmission profile has effects on colors of cooler
stars by making their $B-V$ color marginally bluer.

  \begin{figure}
     \centering
     \resizebox{\hsize}{!}{\includegraphics[angle=270]{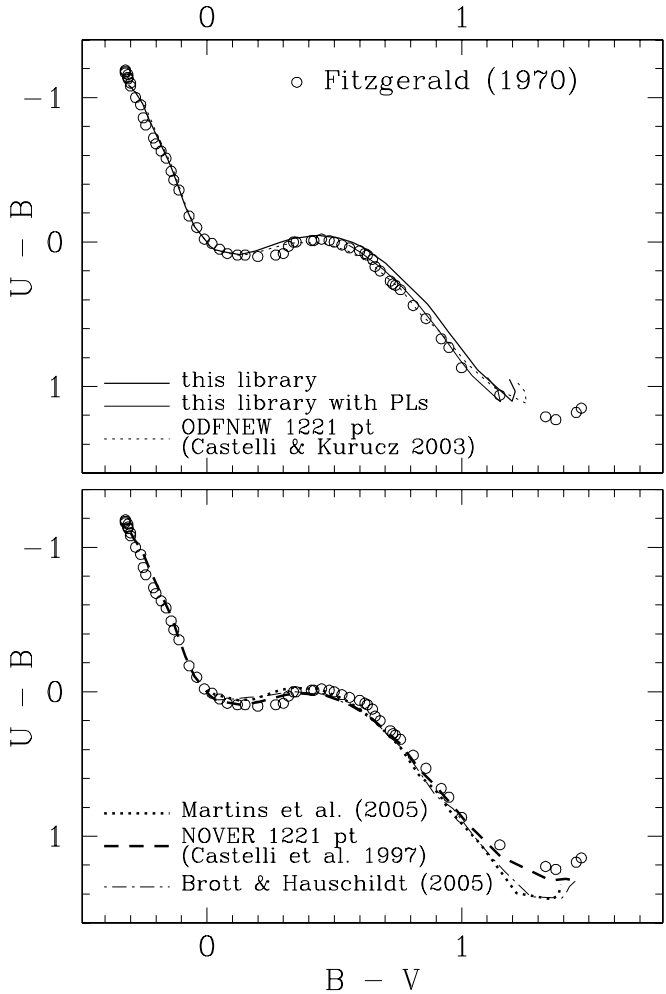}}
     \caption{Comparison of synthetic colors along the Main Sequence
     truncated at M2V. The observed empirical Main Sequence of Fitzgerald
     (1970) serves as a comparison. The adopted band transmission profiles 
     are those of Buser (1978). }
     \label{ubvri1}
  \end{figure}

To assist in the interpretation of the result of the photometric tests we
have also recomputed the spectra of our library that define the MS, this
time with inclusion of PLs (these spectra are available on the library web
page). In Figure~5 and 6 by `NOVER 1221 pt' and `ODFNEW 1221 pt' we mean the
flux distributions that are predicted directly from the model atmosphere
code, not to be confused with the spectra generated by the spectral
synthesis code.

  \begin{figure}
     \centering
     \resizebox{\hsize}{!}{\includegraphics{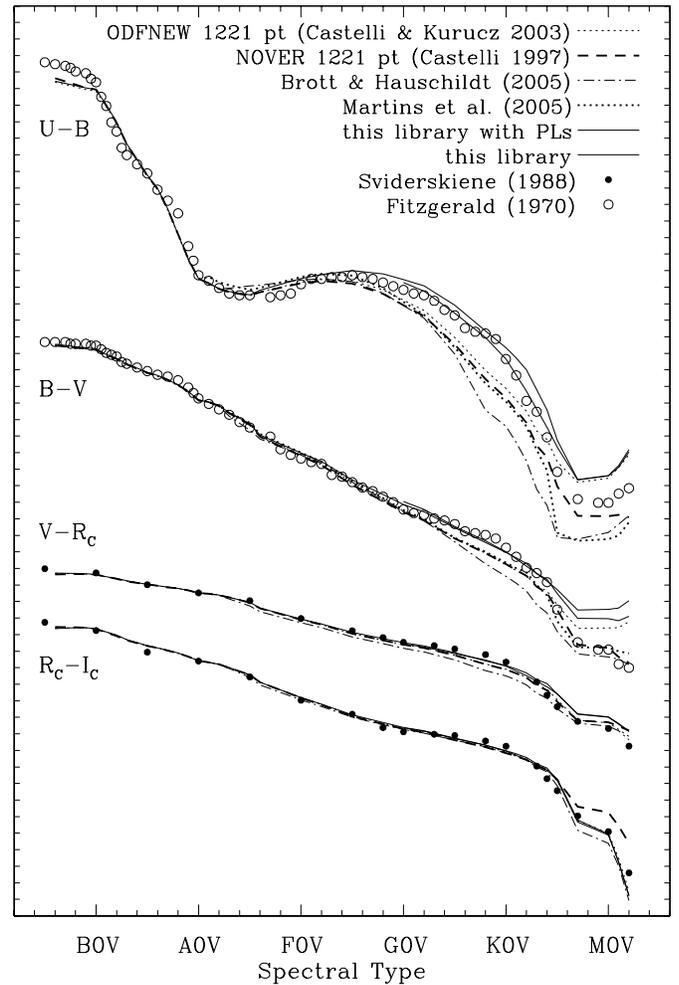}}
     \caption{The figure compares the behaviour of photometric colors along
     the Main Sequence for individual spectral types. The same data of
     Figure ~5 are used. The adopted band transmission profiles for $R_{\rm
     C}$ and $I_{\rm C}$ are those of Bessell (1990).}
     \label{ubvri2}
  \end{figure}

Figure~5 illustrates the behaviour on classical 2-color diagrams. Our
library perform similarly well to the others on the $U-B$,~$B-V$ plane for
colors bluer than 0.7 in $B-V$. For redder colors, the synthetic MS has
$U-B$ colors too blue by some hundredths of a magnitude. The inclusion of
PLs in the computation brings the synthetic MS in close agreement with the
observed one and very close to the ODFNEW 1221 pt flux distributions. The
effect of adopting the ODFNEW models atmospheres affects the lowest
temperatures where the synthetic spectra (with and without PLs, as well as
the 1221 pt flux distributions) fails to match the very red colors observed
in real stars, with the synthetic MS displaying a small upward {\em hook}.

Figure~6 illustrates the data in a different format and expanded scale. The
difference in the single color between real stars and synthetic spectra are
plotted as function of the spectral type along the MS. This diagram
highlights better the scatter between observed colors and those from
synthetic spectra, particularly in $U-B$. It shows that, in all colors,
significant departures of our library from observations are limited to
spectral types cooler than K4V ($T_{\rm eff} <$4500~K).

  \begin{table}
   \centering
   \caption[]{Comparison between different calibrations for colors and temperatures
              of late type main sequence stars.  
              $a$: Bessell  (1990);
              $b$: Bessell  (1995); 
              $c$: Strayzis (1992); 
              $d$: Bertone et al. (2004) using Kurucz spectra;
              $e$: Bertone et al. (2004) using Phoenix spectra; 
              $f$: Drilling and Landolt (2000);
              $g$: Straizys and Kuriliene (1981).
              }
   \includegraphics[width=8.0cm]{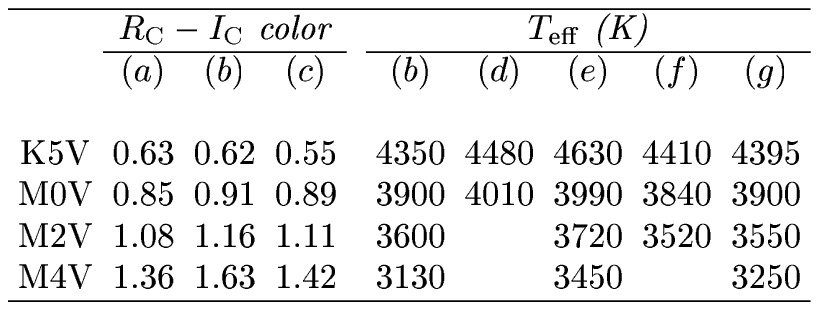}  
   \label{colors}
  \end{table}

Having already proved the satisfactory spectroscopic performance of our
library, we refrain from discussing any further its broad-band photometric
performances and that of the other libraries listed in Table~1, which is
clearly beyond the scope of this paper.  Figures~5 and 6 should not be
considered as the ultimate comparison of the photometric performances of the
available synthetic libraries, even if we placed all libraries
on the same reference system, using the same $U$$B$$V$$R_{\rm C}$$I_{\rm C}$
band profiles and the same $T_{\rm eff}$/$\log g$ conversion to spectral
type for all of them. Even if different reconstruction of the
$U$$B$$V$$R_{\rm C}$$I_{\rm C}$ band profiles or $T_{\rm eff}$/$\log g$
conversion would have been used, the spread among different synthetic
libraries and with respect to observed colors would have remained. As a
matter of the caution that one has to bear in main in such comparisons,
Table~4 shows the widely different $T_{\rm eff}$ calibration that have been
published for the cooler spectral types along the MS considered in Figures~5
and 6.

\begin{acknowledgements}
Support to R.S. by MIUR COFIN2001 program is acknowledged. We would like to thank
S.Ansari for managing the library access via the ESA web portal within the framework
of GAIA support activities and M.Fiorucci for useful support in connection with
treatment of RAVE data. The careful scrutiny and stimulating comments by the anonymous
referee were deeply appreciated. 
\end{acknowledgements}

\newpage

  \begin{figure*}
     \centering
     \resizebox{\hsize}{!}{\includegraphics{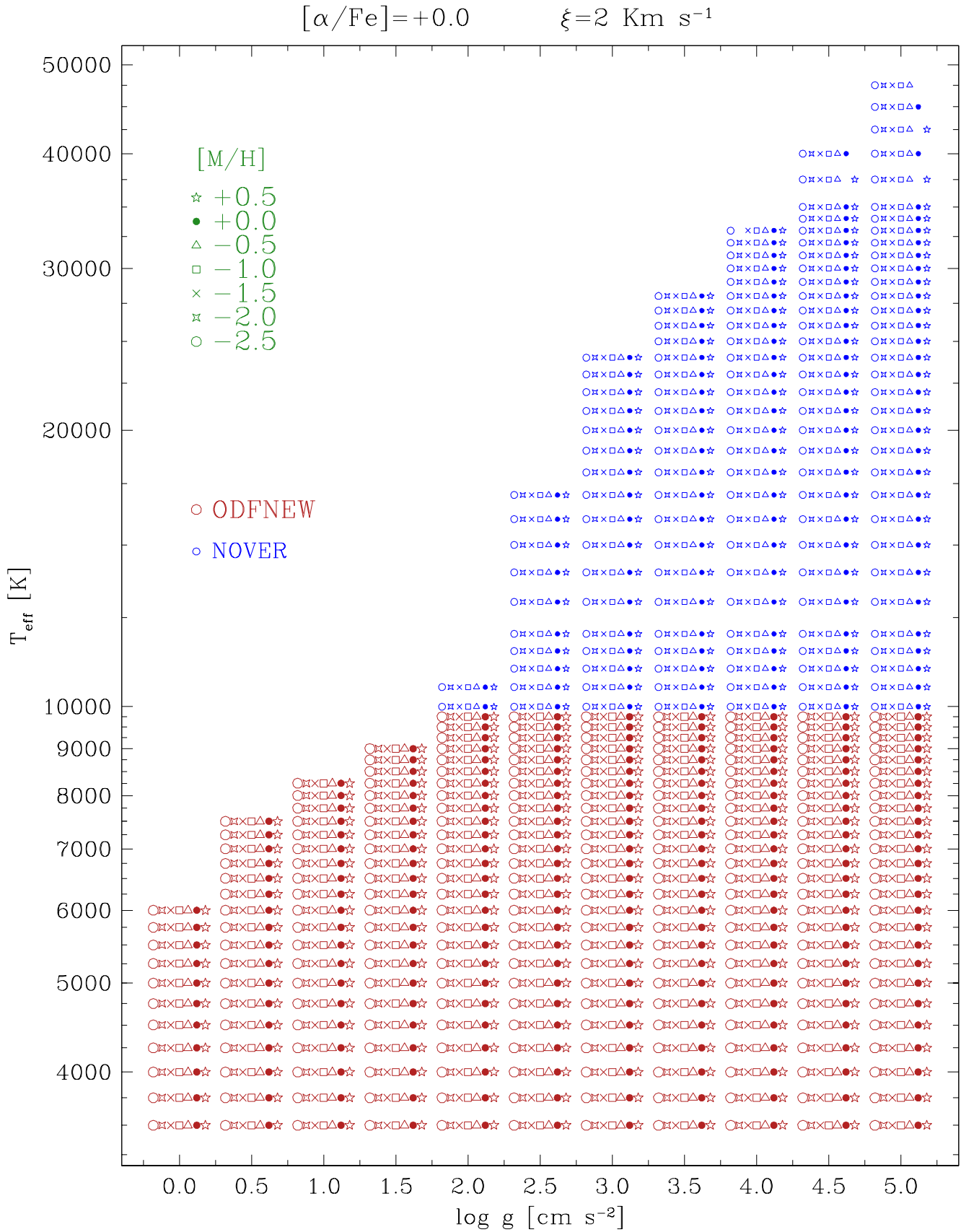}}
     \caption{Figure 1a, to be published electronic only.}
     \label{fig1a}
  \end{figure*}

  \begin{figure*}
     \centering
     \resizebox{\hsize}{!}{\includegraphics{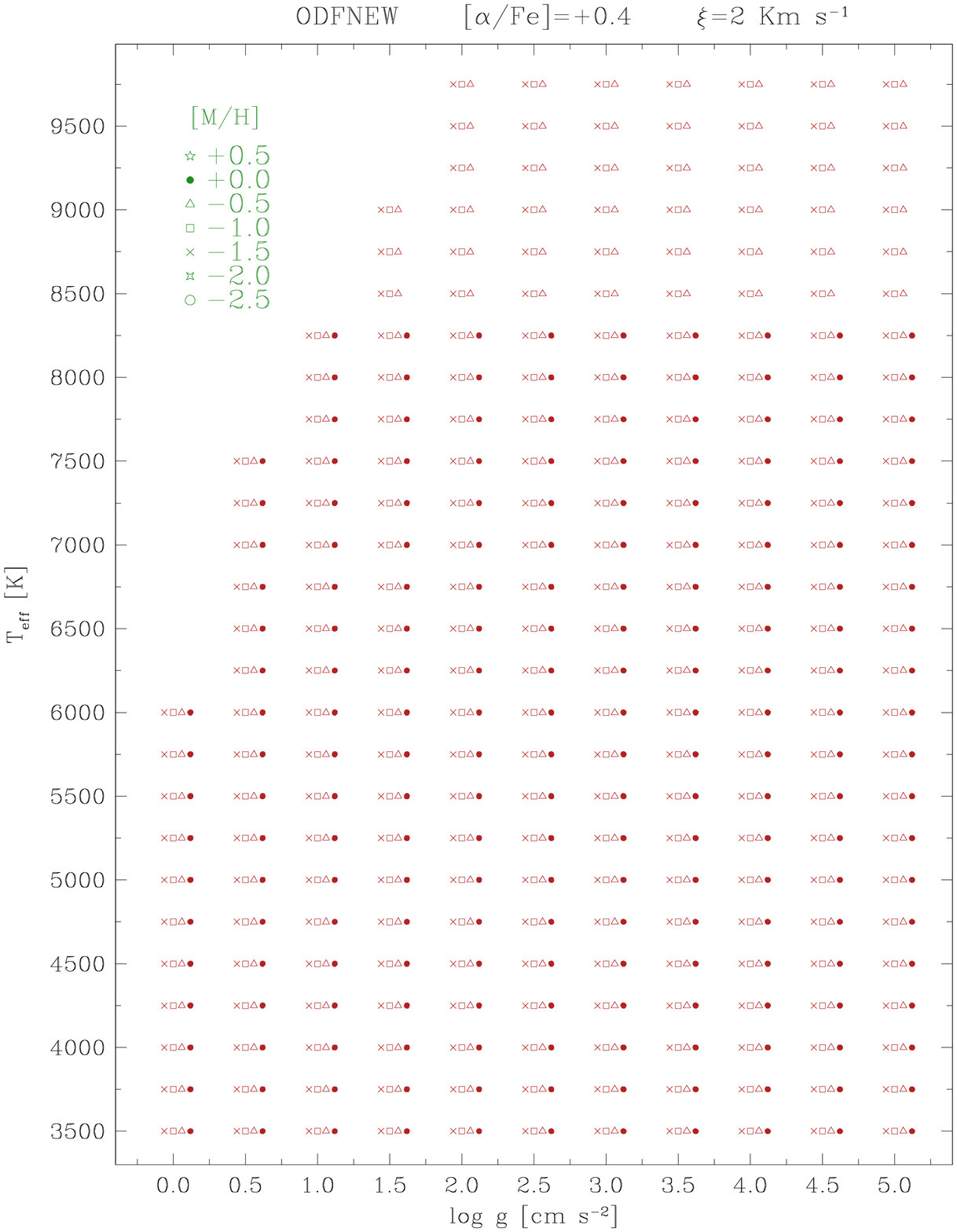}}
     \caption{Figure 1b, to be published electronic only.}
     \label{fig1a}
  \end{figure*}

  \begin{figure*}
     \centering
     \resizebox{\hsize}{!}{\includegraphics{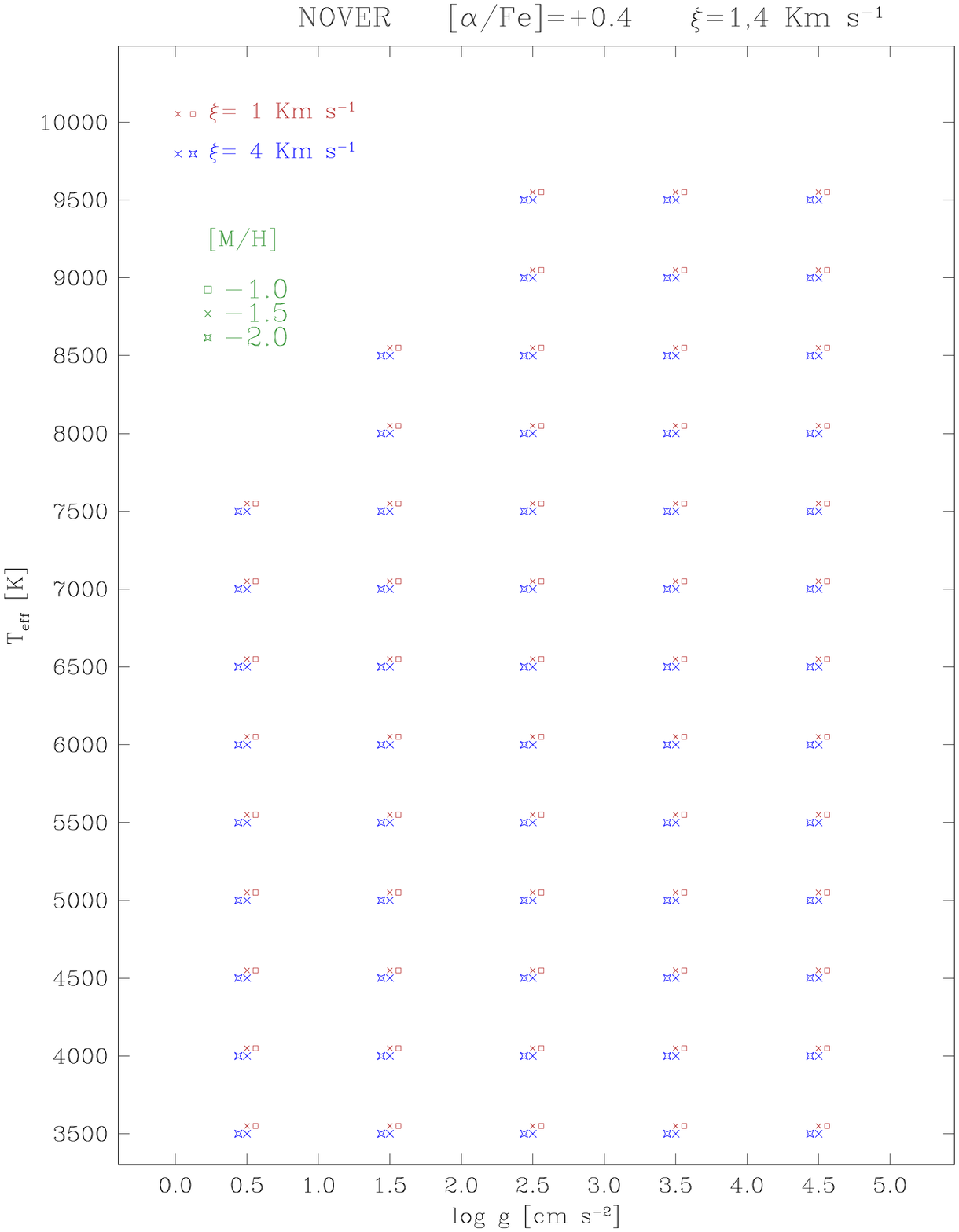}}
     \caption{Figure 1c, to be published electronic only.}
     \label{fig1c}
  \end{figure*}

\end{document}